\documentclass[prX,amsmath,amssymb,superscriptaddress]{revtex4}
\usepackage{amsmath}
\usepackage{latexsym}
\usepackage{graphicx}
\usepackage{color}
\usepackage{bm}

\newcommand{\ket}[1]{|#1\rangle}

\begin{document}

%\title{Three-photon entanglement in energy and time }
\title{Three-photon energy-time entanglement}

\author{L. K. Shalm}
\author{D. R. Hamel}
\author{Z. Yan}

\affiliation{Institute for Quantum Computing and Department of Physics \& Astronomy, University of \\Waterloo, Waterloo, Canada, N2L 3G1}

\author{C. Simon} 
\affiliation{Institute for Quantum Information Science and Department of Physics and Astronomy, University of Calgary, Calgary, Alberta T2N 1N4, Canada}

\author{K. J. Resch}
\author{T. Jennewein}
\affiliation{Institute for Quantum Computing and Department of Physics \& Astronomy, University of \\Waterloo, Waterloo, Canada, N2L 3G1}

\begin{abstract}
Entangled quantum particles have correlations stronger than those allowed by classical physics.  These correlations are the focus of of the deepest issues in quantum mechanics\cite{PhysRev.47.777, Bell1964Original,Aspect:1982p6501} and are the basis of many quantum technologies.  The entanglement of discrete particle properties has been studied extensively in the context of quantum computing \cite{nielsen00}, cryptography \cite{PhysRevLett.67.661}, and quantum repeaters \cite{PhysRevLett.81.5932}
 while entanglement between the continuous properties of particles may play a critical role in improving the sensitivity of gravitational wave detectors\cite{KGoda:2008p5310,Hage:2008fk}, atomic clocks\cite{Ye:2008p5379}, and other high precision instruments.  The attributes of three or more entangled particles are fundamentally different from those of two entangled particles\cite{GHZOriginal,PhysRevD.35.3066,Pan:2000uq, 1367-2630-11-7-073051,PhysRevA.83.062108}. While the discrete variables of up to 14 ions\cite{PhysRevLett.106.130506} and the continuous variables between three intense optical beams\cite{PhysRevLett.91.080404,Coelho:2009cx} have been entangled, it has remained an open challenge to entangle the continuous properties of more than two individual particles.  Here we experimentally demonstrate genuine tripartite continuous-variable entanglement between three separated particles.  In our setup the three particles are photons created directly from a single input photon; the creation process leads to quantum correlations between the colours, or energies, and emission times of the photons.  The entanglement between our three photons is the three-party generalization of the Einstein-Podolsky-Rosen (EPR)\cite{PhysRev.47.777} correlations for continuous variables, and allows for new fundamental tests of quantum mechanics to be carried out.  Our scheme can be extended to carry out multi-particle Franson interferometry\cite{Franson:1989uu,Kwiat:1993p268}, and opens the possibility of using additional degrees of freedom in our photons to simultaneously engineer discrete and continuous-variable hyper-entangled states that could serve as a valuable resource in a wide variety of quantum information tasks.
\end{abstract}

\maketitle

We directly generate three entangled photons using the nonlinear process of cascaded spontaneous parametric downconversion (C-SPDC)\cite{Hubel:gw}.  In downconversion, a pump photon, with frequency $\omega_{p}$, inside a nonlinear material will occasionally fission into a pair of daughter photons with frequencies $\omega_{0}$ and $\omega_{1}$.  The total energy in the process is conserved\cite{PhysRevA.56.1534} with $\hbar \omega_{p}=\hbar \omega_{0}+\hbar \omega_{1}$.   The daughter photons  share strong energy and time correlations that are the hallmark of entanglement\cite{Franson:1989uu,Kwiat:1993p268}.  The SPDC process is repeated with one of these daughter photons, at $\omega_{0}$, now serving as the pump, creating a pair of granddaughter photons simultaneously at $\omega_{2}$ and $\omega_{3}$.  Again energy is conserved, and the total energy of the the three photons created in C-SPDC must sum to the energy of the pump: $\hbar \omega_{p}=\hbar \omega_{1}+\hbar \omega_{2}+\hbar \omega_{3}$.   The simplified representation of our three-photon state in frequency space, assuming a monochromatic pump,  has the form
\begin{equation}
\label{E:3PhotonState}
\Psi_{CSPDC} \approx \int_{\omega_1} \int_{\omega_2} d\omega_1 d\omega_2
G_1(\omega_1,\omega_p-\omega_1) G_2(\omega_2, \omega_p-\omega_1 - \omega_2)
a_1^\dagger(\omega_1) a_2^\dagger(\omega_2) a_3^\dagger(\omega_p-\omega_1 - \omega_2)\ket{0} ,
\end{equation}
where $G_1(\omega_1,\omega_p-\omega_1)$ and $G_2(\omega_2, \omega_p-\omega_1 - \omega_2)$ are the joint-spectral functions resulting from the phase-matching conditions  of the first and second SPDC crystals respectively \cite{PhysRevA.57.2076}.  The three photons, consequently, share strong spectral correlations and exhibit genuine tripartite energy-time entanglement.  

To verify the tripartite entanglement of the photons generated in our C-SPDC process we use continuous variable entanglement criteria, that we derive based on the work of van Loock and Furusawa\cite{vanLoock:2003hn}, for position and momentum.  Consider three separable particles each described by the dimensionless observables  $x_k$, $p_k$ ($k=1,2,3$) fulfilling the commutation relations $[x_k,p_l]=i \delta_{kl}$ (note: van Loock and Furusawa\cite{vanLoock:2003hn} use a different commutation relation than the one used here).  Each individual particle must satisfy the uncertainty relationship $\Delta x_{i} \Delta p_{i} \geq 1/2$. Together, all three particles must satisfy the following position-momentum uncertainty inequalities (see Supplementary Information for details):
\begin{eqnarray}
\label{FurusawaProd1}
\Delta(x_2-x_1) \Delta(p_1+p_2+p_3) \geq 1,\\
\label{FurusawaProd2}
\Delta(x_3-x_2) \Delta(p_1+p_2+p_3) \geq 1,\\
\label{FurusawaProd3}
\Delta(x_3-x_1) \Delta(p_1+p_2+p_3) \geq 1.
\end{eqnarray}
Violating any one of these inequalities is sufficient to demonstrate that a state contains some entanglement. Violating any two inequalities demonstrates that the state is fully inseparable\cite{PhysRevLett.83.3562}. For pure states full inseparability implies genuine tripartite entanglement\cite{Bourennane:2004lr}. However, full inseparability and genuine tripartite entanglement are not, in general, the same thing. Mixtures of bipartite entangled states that are fully inseparable but not genuinely tripartite entangled are also capable of violating two of the above inequalities.  A more general entanglement criterion is therefore required to detect genuine tripartite entanglement. In the Supplementary Information we provide an overview of the definitions of full inseparability and genuine tripartite entanglement and derive the following inequalities:
\begin{eqnarray}
\label{sum1}
\left[ \Delta(x_2-x_1) +\Delta(x_3-x_1) \right] \Delta(p_1+p_2+p_3) &\geq& 1. \\
\label{sum2}
\left[ \Delta(x_2-x_1) +\Delta(x_3-x_2) \right] \Delta(p_1+p_2+p_3) &\geq& 1,\\
\label{sum3}
\left[\Delta(x_3-x_2) +\Delta(x_3-x_1) \right] \Delta(p_1+p_2+p_3) &\geq& 1, \\
\label{sumall3}
\left[ \Delta(x_2-x_1)  +\Delta(x_3-x_1) + \Delta(x_3-x_2)\right]\Delta(p_1+p_2+p_3) &\geq& 2.
\end{eqnarray}
Violating any one of them is sufficient to demonstrate genuine tripartite entanglement.

The position and momentum operators $x$ and $p$ are well-defined for narrow-band photons\cite{Fedorov:2005fh}, such as those generated by our C-SPDC process, with the usual commutation relation $[x,p]=i$.  Because photons propagate at the speed of light, $c$, measuring the arrival time, $t$, of a photon at a single-photon detector is  equivalent to measuring its longitudinal position $x$ ($t=x/c$), and measuring its frequency, $\omega$, is equivalent to measuring its longitudinal momentum $p$ ($\hbar \omega = cp$). Using this correspondence it is possible to write down the energy-time equivalents to the inequalities in equations \ref{sum1}-\ref{sumall3}:
\begin{eqnarray}
\label{MinUncertaintyRelations1}
\left[ \Delta(t_{2}-t_{1}) +\Delta(t_{3}-t_{1}) \right] \Delta(\omega_{1}+\omega_{2}+\omega_{3}) &\geq& 1. \\
\label{MinUncertaintyRelations2}
\left[ \Delta(t_{2}-t_{1}) +\Delta(t_{3}-t_{2})\right] \Delta(\omega_{1}+\omega_{2}+\omega_{3}) &\geq& 1, \\
\label{MinUncertaintyRelations3}
\left[\Delta(t_{3}-t_{2}) +\Delta(t_{3}-t_{1}) \right] \Delta(\omega_{1}+\omega_{2}+\omega_{3}) &\geq& 1, \\
\label{MinUncertaintyRelations4}
\left[ \Delta(t_{2}-t_{1})  +\Delta(t_{3}-t_{1}) + \Delta(t_{3}-t_{2})\right]\Delta(\omega_{1}+\omega_{2}+\omega_{3}) &\geq& 2.
\end{eqnarray}

States of the form in Equation \ref{E:3PhotonState}  can violate all the inequalities maximally, ie the left-hand side goes to zero, and thus exhibit genuine tripartite entanglement. 

Measuring the difference in arrival times of the three photons using fast single-photon detectors gives the required timing uncertainties for testing the inequalities. However, directly measuring the frequencies of each individual photon with the precision needed (sub-GHz resolution over a bandwidth of several THz) to violate the inequalities is not  technologically feasible. A solution to this problem is to use the fact that energy is conserved in the process of downconversion.  The energy of the pump is equal to the energy of the three daughter photons created in C-SPDC ($\hbar \omega_{p}= \hbar \omega_{1}+\hbar \omega_{2}+\hbar \omega_{3}$); measuring the frequency of the pump provides a direct measurement of the total frequency, of the three daughter photons required by the inequalities.  This conservation of frequency has previously been tested using Franson interferometry\cite{Franson:1989uu,Kwiat:1993p268}, and has been measured down to a line width of 200 kHz, an uncertainty much smaller than the scale considered in our experiment, in second-harmonic generation\cite{Sun:2007kv,Ikegami199669} (the time-reversed process of SPDC). 

To create our three entangled photons using C-SPDC (see  Figure 1), first a narrowband pump laser at 404 nm is used to produce a pair of non-degenerate SPDC photons at 776 nm and 842 nm.  The photon at 776 nm is then sent through a second SPDC crystal where a pair of granddaughter photons at 1530 nm and 1570 nm are generated (see Methods Summary for more details).  This process leaves the 842 nm, 1530 nm and 1570 nm photons entangled in energy and time.  Our setup detects an average of 7 triples/hour, from which we can infer the generation of 45 triples/minute accounting for losses due to coupling and detection. In order to obtain sufficient photon counts with small statistical fluctuations, data was collected for a total of 72.6 hours. 
%%%%%%%%%%%%%%%%
%Experiment Figure
\begin{figure}
\centering
\includegraphics[width=6in]{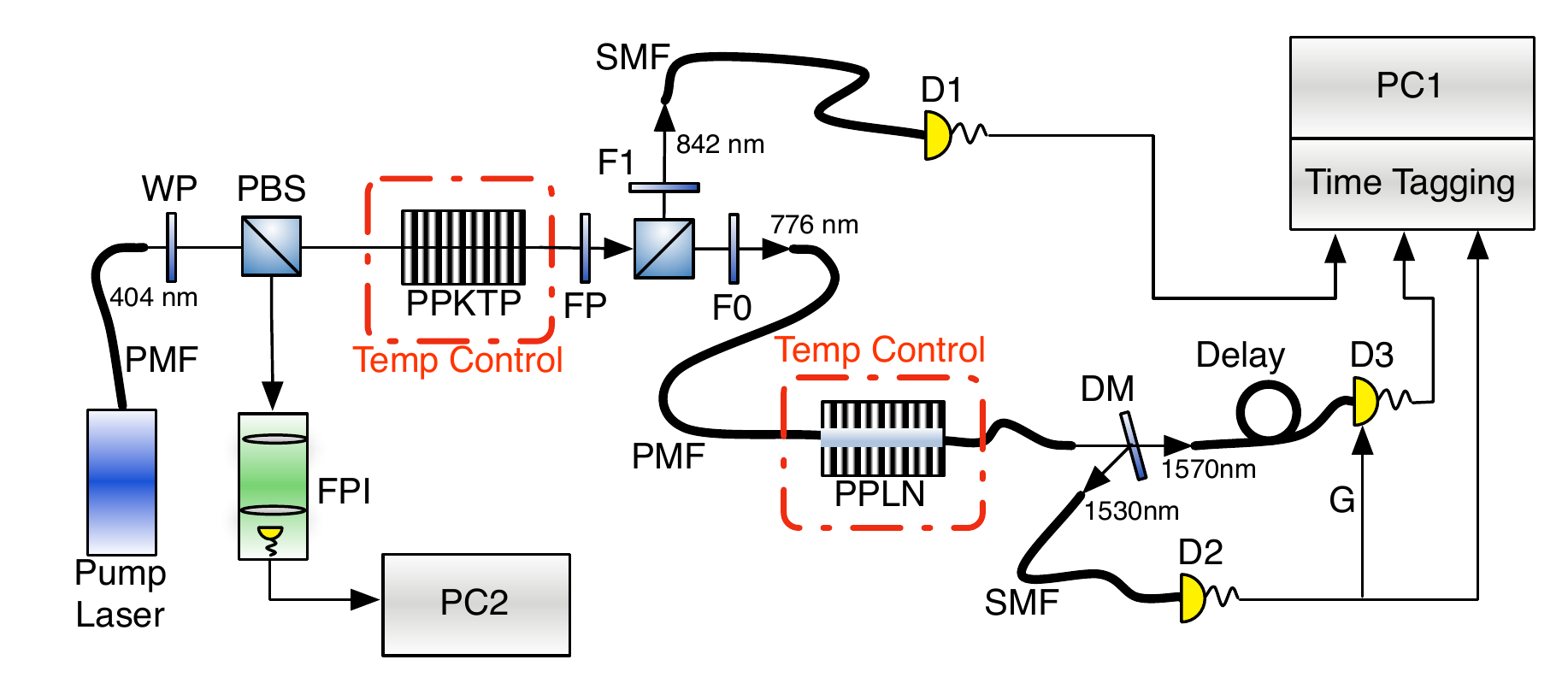}
\caption{Our three entangled photons are created using C-SPDC.  A narrowband pump laser at 404 nm downconverts into a pair of orthogonally polarised photons at 842 nm and 776 nm inside a periodically-poled Potassium Titanyl Phosphate (PPKTP) crystal.  A filter (FP) removes the remaining pump light.  A polarising beamsplitter is used to separate the two photons, and narrowband filters, F0 and F1, are used to block stray light.  The photon at 842 nm is coupled into a single-mode fibre and sent to the single-photon detector D1.   The photon at 776 nm is coupled into single mode fibre and sent to a periodically-poled Lithium Niobate (PPLN) waveguide where it downconverts into a pair of photons at 1530 nm and 1570 nm.  The photons are outcoupled into free space where a dichroic mirror is used to split the photons.  The photons are then coupled back into single-mode fibre and sent to single-photon detectors D2 and D3 (see Methods Summary for more information about the detectors). The signals from all three detectors are sent to a time tagging unit, and a computer (PC1) is used to process coincidence events.  The spectrum of the 404 nm pump laser is continuously monitored throughout the run using a Fabry-Perot interferometer (FPI) controlled by a second computer (PC2). }
\label{F:Setup}
\end{figure}

%%%%%%%%%
%3D Triples Plot
\begin{figure}
\centering
\includegraphics[width=6in]{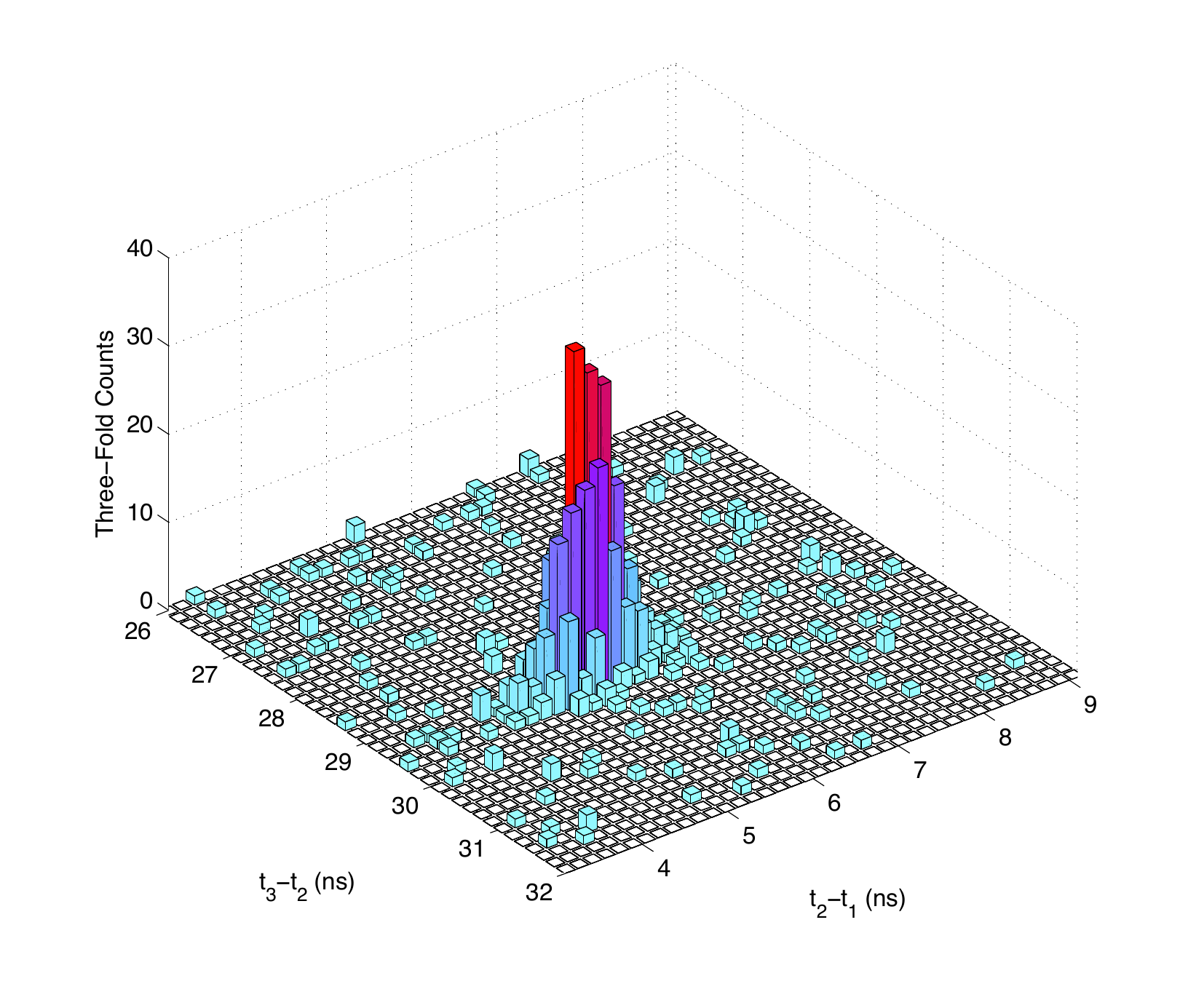}
\caption{2D histogram of the timing information for the measured triple coincidences over 72.6 hours.  The triple events are all localized to a small region of the histogram, indicating strong correlations in the arrival times of the three photons.} 
\label{F:3D}
\end{figure}

The timing information from the detections was analyzed, and the triple coincidence counts binned into a 2D histogram based on $t_{2}-t_{1}$ and $t_{3}-t_{2}$ as shown in Figure 2.  From the histogram it is clear that the photon arrivals are tightly correlated in time.  The uncertainty in the arrival time between any pair of photons can be found by integrating over the arrival time of the other photon, removing its dependence as shown in Figure 3.  From these integrated histograms we find that $\Delta(t_{2}-t_{1})=0.37\pm0.02$ ns, $\Delta(t_{3}-t_{2})=0.162\pm0.004$ ns, and $\Delta(t_{3}-t_{1})=0.31\pm0.02$ ns.  Our measurements are limited by the timing jitter in our detectors and the resolution of our time-tagging unit (156 ps).  The effect of the jitter can be clearly seen in the elliptical shape of the 2D arrival time histogram---the jitter on the detector used to detect the 842 nm photon is a factor of two larger than the jitter of the two telecom detectors.  This is reflected in the uncertainty $\Delta(t_{3}-t_{2})$ which is approximately a factor of two smaller than either $\Delta(t_{2}-t_{1})$ or $\Delta(t_{3}-t_{1})$. Alternatively we can study the two-photon coincidences between detectors D1 \& D2 and D2 \& D3  independent of the third detector (see Supplementary Information) to verify that integrating over the third photon yields the correct two-photon timing histograms.  The need for gating with the 1570 nm detector ($t_{3}$) prevents the coincidences between $t_{3}$ and $t_{1}$ from being analyzed independent of $t_{2}$, but the 50 ns gate width is much larger than the uncertainty in the arrival time of a photon, and approximates the response of a free-running detector.  

%%%%%%%%%%
%2 coincidence data
\begin{figure}
\centering
\includegraphics[width=3in]{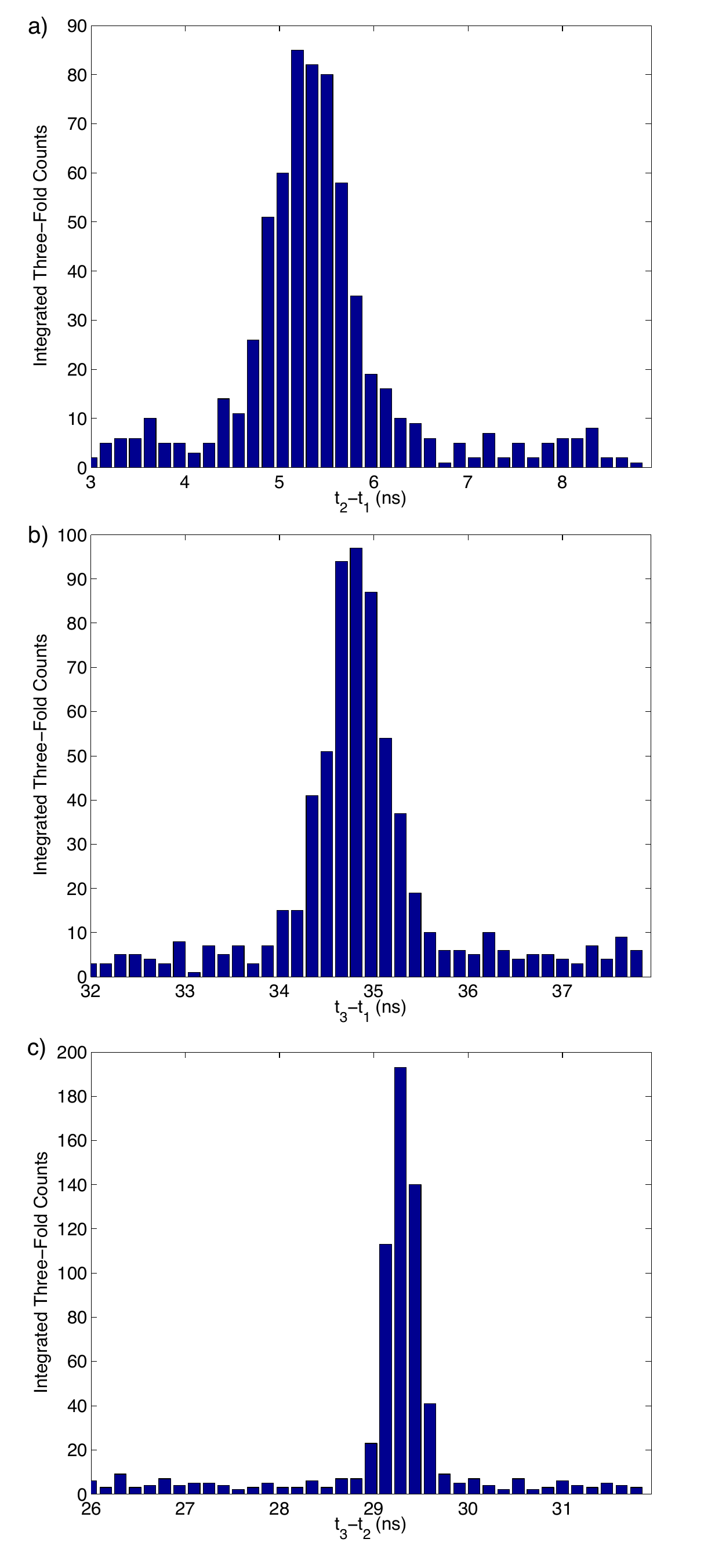}
\caption{Histograms of the difference in arrival times between two of the three photons measured over 72.6 hours.  Each histogram is obtained by integrating the triples counts over the arrival time of the third photon to remove its dependence on the results.  From this we find the uncertainty in the arrival times of the photons to be a) $\Delta(t_{2}-t_{1})=0.37\pm0.02$ ns, b) $\Delta(t_{3}-t_{2})=0.162\pm0.004$ ns, and c) $\Delta(t_{3}-t_{1})=0.31\pm0.02$ ns.  The timing uncertainties were verified using two-fold coincidence data that was obtained at the same time as the three-fold coincidence data (see Supplementary Information).}
\label{F:2CoincData}
\end{figure}

The energy uncertainty of the photon triplets is given by the energy uncertainty in the 404 nm pump photons.  To measure the uncertainty in the pump energy, a scanning Fabry-Perot interferometer (FPI) was used to continuously monitor the bandwidth of the 404 nm laser throughout the experiment. Due to instabilities caused by temperature fluctuations, the measured bandwidth fluctuates over time as shown in Figure 4a, leading to the distribution in Figure 4b.  The average value and standard deviation of this distribution yield a pump bandwidth of $\Delta \omega_{p} /2\pi=(6 \pm 2)$ MHz.  

\begin{figure}
\centering
\includegraphics[width=6in]{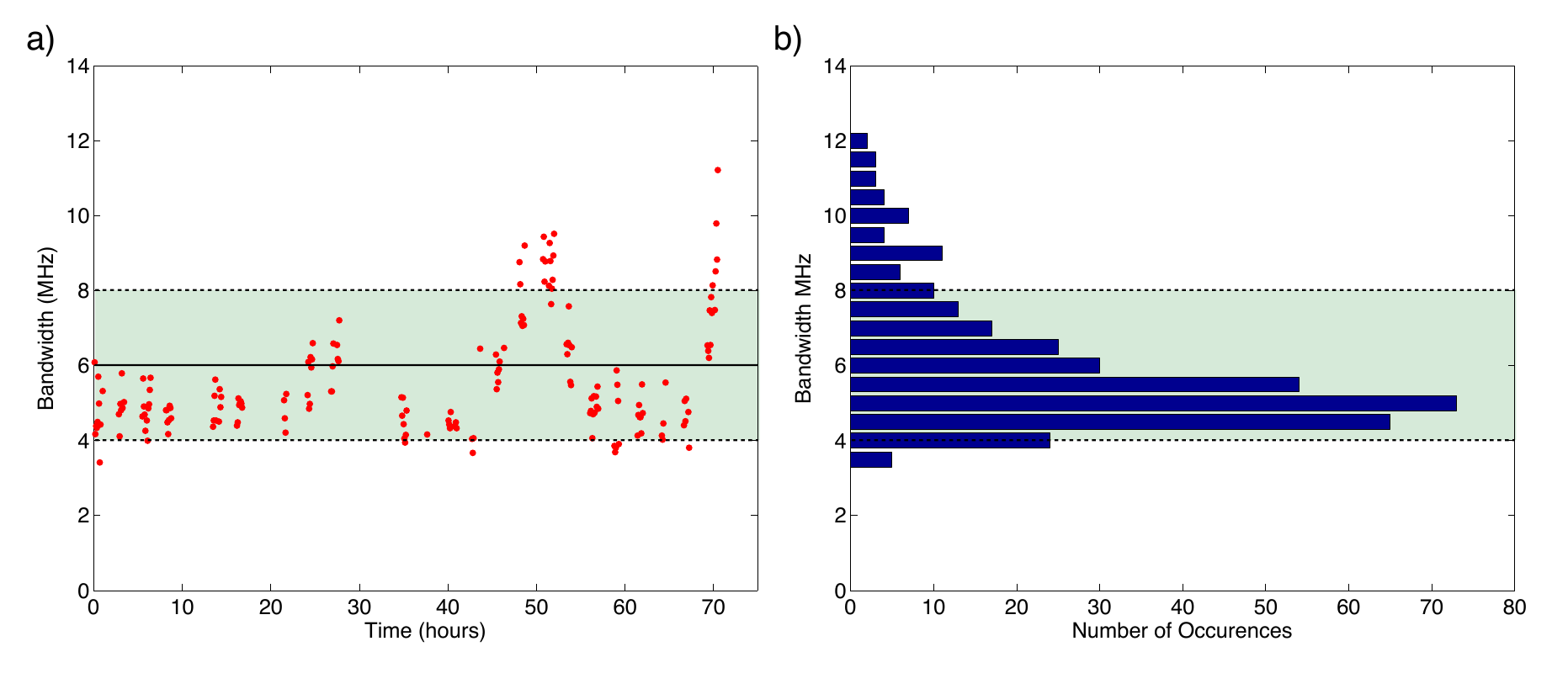}
\caption{Bandwidth of the pump photons as measured by a Fabry-Perot interferometer at five minute intervals during the 72.6 hour run.   a) The measured bandwidth as a function of time. The fluctuations in the measured bandwidths are the result of thermal drifts in the apparatus over the course of the run.  b) Histogram of the measured pump bandwidths over the duration of the run.  The average bandwidth is measured to be $\Delta \omega_{p}/2\pi=$6 MHz with a standard deviation of 2 MHz (illustrated by the shaded regions in the graphs).}
\label{F:FP}
\end{figure}

The four measured time-bandwidth products for our three photons are
\begin{eqnarray}
\label{MinUncertaintyRelationsResult1}
\left[ \Delta(t_{2}-t_{1}) +\Delta(t_{3}-t_{1}) \right] \Delta(\omega_{1}+\omega_{2}+\omega_{3}) &=& 0.03 \pm 0.01,\\
\label{MinUncertaintyRelationsResult2}
\left[ \Delta(t_{2}-t_{1}) +\Delta(t_{3}-t_{2})\right] \Delta(\omega_{1}+\omega_{2}+\omega_{3})  &=& 0.02 \pm 0.01, \\
\label{MinUncertaintyRelationsResult3}
\left[\Delta(t_{3}-t_{2}) +\Delta(t_{3}-t_{1}) \right] \Delta(\omega_{1}+\omega_{2}+\omega_{3}) &=& 0.018 \pm 0.005, \\
\label{MinUncertaintyRelationsResult4}
\left[ \Delta(t_{2}-t_{1})  +\Delta(t_{3}-t_{1}) + \Delta(t_{3}-t_{2})\right]\Delta(\omega_{1}+\omega_{2}+\omega_{3})  &=&0.03 \pm 0.01.
\end{eqnarray}

Our three photons strongly violate inequalities \ref{MinUncertaintyRelations1}-\ref{MinUncertaintyRelations4}  and are genuinely tripartite entangled.  Our state exhibits energy-time correlations close to to the ideal state described in Equation \ref{E:3PhotonState} where the time-bandwidth products are exactly zero.  This state is the continuous-variable analogue to the famous Greenberger, Horne and Zeilinger (GHZ) entangled state\cite{GHZOriginal,vanLoock:2003hn}, and the natural extension of the two-party continuous-variable EPR state \cite{Franson:1989uu,Kwiat:1993p268,JohnCHowell:2004fc}. 

Recent improvements in telecom wavelength detectors \cite{Collins2011} and advances in nonlinear materials promise to dramatically increase our detected triples rate \cite{Corona:11,Richard:11}.  Furthermore, new techniques to enhance the strength of nonlinear effects \cite{PhysRevLett.106.120403, Langford:2011fk} means that our scheme can in principle be scaled up to larger photon numbers.  A major advantage of our states is that the continuous-variable entanglement is distributed amongst three individual photons, each at a different, tuneable, wavelength, enabling the creation of hyper-entangled states that simultaneously take advantage of both discrete and continuous  variable quantum correlations.  This multiplexing of entanglement over multiple discrete and continuous degrees of freedom may have important applications in quantum communication tasks. For example, a slight modification to our setup would enable a photon at 776 nm to be interfaced with an atomic storage medium like Rb while the remaining two photons are transmitted over telecom fibres to remote quantum nodes.  This would open up new possibilities in the storage and distribution of quantum information needed for quantum computing, cryptography, and secret sharing, and could lead to new fundamental tests of quantum mechanics.  

\clearpage

\subsection{Methods Summary}
$\\$
In our setup (shown if Figure 1) we use a grating-stabilized pump laser with a wavelength of 404 nm and a bandwidth of 5 MHz (Toptica Bluemode) to pump a 30 mm PPKTP crystal phase matched for Type-II SPDC.  A pair of orthogonally polarised signal and idler photons at 842 nm and 776 nm respectively are generated co-linearly, and a polarising beam splitter (PBS) is used to separate them.  The signal photon at 842 nm is coupled into an optical fibre and sent to a Si single-photon detector.  With 12 mW of pump power $10^{6}$ signal photons/s are detected.  The idler photon is fibre-coupled and sent to a second SPDC crystal, a 30 mm Type-I phase matched PPLN waveguide (HC Photonics), where it fissions into a pair of granddaughter photons at 1570 nm and 1530 nm.  These granddaughter photons are outcoupled into free space and then split using a dichroic mirror.   The photon at 1530 nm is sent to a free-running InGaAs/InP-Avalanche Photo Diodes (Princeton Lightwave, Negative Feedback Avalanche Diode - NFAD) detector cooled to 193 K operating at 10\% efficiency with approximately 100 dark counts/s.  This detector is used to gate a second InGaAs/InP detector (iD Quantique, id201-SMF-ULN) operating at 25\% detection efficiency with a 50 ns gate window to detect the granddaughter photon at 1570 nm.  The gated detector had a much higher dark count rate of approximately $5\times10^{-5}$ dark counts/(ns of gate).  The arrival times of each of photons in the three detectors are recorded by a time-tagging system (DotFast/UQDevices) with 156 ps resolution.  In this way all the timing statistics from the two-fold and three-fold coincidence events generated by the C-SPDC process can be measured.
%\end{methods}

\subsection{Acknowledgements}
 We thank the Ontario Ministry of Research and Innovation ERA, QuantumWorks, NSERC, OCE, Industry Canada, Canadian Institute for Advanced Research (CIFAR) and CFI for financial support.  C.S. acknowledges  support by an Alberta Innovates Technology Futures (AITF) New Faculty Award. We thank B. Boulanger  and A. M. Steinberg for for useful discussions.
\subsection{Competing Interests} The authors declare that they have no
competing financial interests.
\subsection{Author Contributions} L.K.S. and D.R.H. carried out the experiment.  C.S., K.J.R. and T.J. conceived of the experiment.  Z.Y. developed the detectors and electronics used in the experiment.  Data was analysed by L.K.S., D.R.H., K.J.R., and T.J.  All authors contributed to the writing of the manuscript.
 \subsection{Correspondence} Correspondence and requests for materials should be addressed\\ to L.K.S.~(email: kshalm@uwaterloo.ca) and T.J. (email: tjennewe@uwaterloo.ca).

\bibliography{Energy-time}

\newpage

\section{Supplementary Information}
\subsection{Fully inseparable versus genuine tripartite entanglement}

There are two definitions that have been introduced for
describing three-particle entanglement: \emph{fully
inseparable} \cite{PhysRevLett.83.3562} and \emph{genuine tripartite entanglement}
\cite{Bourennane:2004lr}. While these terms are sometimes used interchangeably in
the literature, they mean different things. In
\cite{PhysRevLett.83.3562} a classification scheme for three
particle entanglement was developed based on the following set
of tripartite states:
\begin{eqnarray}
\label{Insep1}
\rho&=&\sum_{i}\eta_{i} \rho_{1,2,i}\otimes \rho_{3,i},\\
\label{Insep2}
\rho&=&\sum_{i}\eta_{i} \rho_{1,3,i}\otimes \rho_{2,i},\\
\label{Insep3}
\rho&=&\sum_{i}\eta_{i} \rho_{2,3,i}\otimes \rho_{1,i},\\
\label{Insep4}
\rho&=&\sum_{i}\eta_{i} \rho_{1,i}\otimes \rho_{2,i} \otimes \rho_{3,i}.
\end{eqnarray}
The first three states are biseparable as they are
factorizable across a single cut while the final state is said to
be fully separable as all three particles can be factorized.
Tripartite states that cannot be written in \emph{any} of these
forms were defined to be fully inseparable \cite{PhysRevLett.83.3562}.

Genuine tripartite entanglement \cite{Bourennane:2004lr} refers to
states that cannot be written as a convex sum of just fully
separable and biseparable states.  In other words, they cannot
be written as a convex sum of the states in equations \ref{Insep1}-\ref{Insep2}. These two definitions are not equivalent. For example, consider the state,
\begin{eqnarray}
\rho=\frac{1}{2}\rho_{12}\otimes\rho_3 + \frac{1}{2}\rho_1\otimes \rho_{23}.
\end{eqnarray}
This mixed state is not genuine tripartite entangled as it
is a convex sum of biseparable states, yet it may be
fully inseparable.  In section D we present a more explicit example of a state that is fully inseparable but does not contain genuine tripartite entanglement. 

In \cite{vanLoock:2003hn} a criterion for detecting tripartite full inseparability in continuous systems is developed. In the following sections we build on this work to derive a stronger criteria capable of detecting genuine tripartite entanglement.  

\subsection{Uncertainty relations}

Consider three separable particles each described by the dimensionless observables  $x_k$, $p_k$ ($k=1,2,3$) fulfilling the commutation relations $[x_k,p_l]=i \delta_{kl}$ (note: the derivation in \cite{vanLoock:2003hn} uses a commutator with a different scaling factor).  Each individual particle must satisfy the uncertainty relation $\Delta x_{i} \Delta p_{i} \geq 1/2$.  

For a general three particle state $\rho=\sum_{i} \eta_{i} \rho_{123,i}$, the variance in $\Delta^{2}(x_{1}-x_{2})_{\rho}$ is given as:
\begin{eqnarray}
\label{UncertaintyDifference}
\Delta^{2}(x_{1}-x_{2})_{\rho}&=&\left< (x_{1}-x_{2})^{2} \right>_{\rho}-\left<x_{1}-x_{2}\right>^{2}_{\rho}\\
&=& \sum_{i} \eta_{i} \left< (x_{1}-x_{2})^{2} \right>_{i}-\left( \sum_{i} \eta_{i} \left<x_{1}-x_{2}\right>_{i} \right)^{2}\\
&=& \sum_{i} \eta_{i} \left< (x_{1}-x_{2})^{2} \right>_{i}-\left( \sum_{i} \eta_{i} \left<x_{1}-x_{2}\right>_{i} \right)^{2} +\sum_{i} \eta_{i} \left<x_{1}-x_{2}\right>_{i}^{2} -\sum_{i} \eta_{i} \left<x_{1}-x_{2}\right>_{i}^{2} \\
\label{UncDiff2}
&=& \sum_{i} \eta_{i} \Delta^{2}(x_{1}-x_{2})_{i}+\sum_{i} \eta_{i} \left<x_{1}-x_{2}\right>_{i}^{2}-\left( \sum_{i} \eta_{i} \left<x_{1}-x_{2}\right> _{i}\right)^{2}.
\end{eqnarray}
From the Cauchy-Schwarz inequality $\sum_{i} \eta_{i} \left<x_{1}-x_{2}\right>_{i}^{2}\geq\left( \sum_{i} \eta_{i} \left<x_{1}-x_{2}\right>_{i} \right)^{2}$, therefore the last two terms in equation \ref{UncDiff2} will always be greater than or equal to zero.  This leads to the inequality:
\begin{eqnarray}
\label{UncDiff3}
\Delta^{2}(x_{1}-x_{2})_{\rho}&\geq& \sum_{i} \eta_{i} \Delta^{2}(x_{1}-x_{2})_{i}.
\end{eqnarray}
This result holds for any sum or difference of operators of this form. The uncertainty in $(x_{1}-x_{2})$ for the mixed state $\rho$ is greater than or equal to the uncertainties in $(x_{1}-x_{2})$ of the components of the mixture.
\begin{eqnarray}
\label{UncDiff4}
\Delta^{2}(x_{1}-x_{3})_{\rho}&\geq& \sum_{i} \eta_{i} \Delta^{2}(x_{1}-x_{3})_{i},\\
\label{UncDiff5}
\Delta^{2}(x_{2}-x_{3})_{\rho}&\geq& \sum_{i} \eta_{i} \Delta^{2}(x_{2}-x_{3})_{i},\\
\label{UncDiffP}
\Delta^{2}(p_{1}+p_{2}+p_{3})_{\rho}&\geq& \sum_{i} \eta_{i} \Delta^{2}(p_{1}+p_{2}+p_{3})_{i}.
\end{eqnarray}

These results lead to the following position and momentum inequality:
\begin{eqnarray}
\label{MinUncBeforeCZ}
\Delta^{2}(x_1-x_2) \Delta^{2}(p_1+p_2+p_3) &\geq&  \left(\sum_{i} \eta_{i} \Delta^{2}(x_{1}-x_{2})_{i}\right)  \left(\sum_{j} \eta_{j}  \Delta^{2}(p_{1}+p_{2}+p_{3})_{j}\right).
\end{eqnarray}

Applying the Cauchy-Scharwz inequality to the right hand side of the previous equation yields:
\begin{eqnarray}
\label{MinUncAfterCZ1}
\Delta^{2}(x_1-x_2) \Delta^{2}(p_1+p_2+p_3) &\geq&  \left(\sum_{i} \eta_{i} \Delta (x_{1}-x_{2})_{i} \Delta (p_{1}+p_{2}+p_{3})_{i}\right)^{2},\\
\label{MinUncCZ1}
\Delta (x_1-x_2) \Delta (p_1+p_2+p_3) &\geq&  \sum_{i} \eta_{i} \Delta (x_{1}-x_{2})_{i} \Delta (p_{1}+p_{2}+p_{3})_{i}.
\end{eqnarray}

Following the same steps we also find:
\begin{eqnarray}
\label{MinUncCZ2}
\Delta (x_2-x_3) \Delta (p_1+p_2+p_3) &\geq&  \sum_{i} \eta_{i} \Delta (x_{2}-x_{3})_{i} \Delta (p_{1}+p_{2}+p_{3})_{i},\\
\label{MinUncCZ3}
\Delta (x_1-x_3) \Delta (p_1+p_2+p_3) &\geq&  \sum_{i} \eta_{i} \Delta (x_{1}-x_{3})_{i} \Delta (p_{1}+p_{2}+p_{3})_{i}.
\end{eqnarray}

\subsection{Uncertainty relations for detecting continuous variable genuine tripartite entanglement}

In \cite{vanLoock:2003hn} van Loock and Furusawa study the permutations of a particular class of three-particle states defined as:
\begin{eqnarray}
\label{FurusawaState}
\rho=\sum_{i} \eta_{i}\rho_{i,12} \otimes \rho_{i,3},
\end{eqnarray}
where particles $1$ and $2$ can be entangled with one another, but are separable from particle $3$, and derive a set of inequalities: 
\begin{eqnarray}
\label{FurusawaFull1}
[\Delta(x_1-x_2)]^2 +[\Delta(p_1+p_2+p_3)]^2  \geq 2, \\
\label{FurusawaFull2}
[\Delta(x_2-x_3)]^2 +[\Delta(p_1+p_2+p_3)]^2  \geq 2, \\
\label{FurusawaFull3}
[\Delta(x_1-x_3)]^2 +[\Delta(p_1+p_2+p_3)]^2  \geq 2. 
\end{eqnarray}

It is important to note that in \cite{vanLoock:2003hn} a different scaling factor for the commutator is used which leads to the right-hand side of these inequalities having a value of one instead of two in that work. 

As the derivation of inequalities \ref{FurusawaFull1}-\ref{FurusawaFull3} depends only on the form of the commutator, they also hold for rescaled variables $x'_k=s x_k$, $p'_k=p_k/s$ for any scaling factor $s$. Minimizing the expressions on the left of the inequalities with respect to the scaling factor $s$ yields the following product inequalities:

\begin{eqnarray}
\label{FurusawaProd1}
\Delta (x_1-x_2) \Delta (p_1+p_2+p_3) \geq 1,\\
\label{FurusawaProd2}
\Delta (x_2-x_3) \Delta (p_1+p_2+p_3) \geq 1,\\
\label{FurusawaProd3}
\Delta (x_1-x_3) \Delta (p_1+p_2+p_3) \geq 1.
\end{eqnarray}
States that simultaneously violate any two of these inequalities are fully inseparable.  Now consider the following $\rho$ that is a mixture of all possible biseparable and fully separable states. 
\begin{eqnarray}
\label{GeneralMixedState}
\rho=\sum_{i} \eta_{i} \rho_{i,12} \otimes \rho_{i,3}+\sum_{j} \mu_{j}\rho_{j,13} \otimes \rho_{j,2}+\sum_{k} \gamma_{k}\rho_{k,23} \otimes \rho_{k,1}+\sum_{l} \nu_{l}\rho_{l,1}\otimes \rho_{l,2} \otimes \rho_{l,3},
\end{eqnarray} 
where $\eta_{i}$, $\mu_{j}$, $\gamma_{k}$, and $\nu_{l}$ are probabilities with $\sum_{i}\eta_{i}+\sum_{j}\mu_{j}+\sum_{k}\gamma_{k}+\sum_{l}\nu_{l}=1$. With the correct choice of parameters it is possible for $\rho$ to violate two of the van Loock and Furusawa inequalities simultaneously despite not containing \emph{any} genuine tripartite entanglement (an explicit example is given in section D). A more stringent test is needed to look for genuine tripartite entanglement. 

Using the results from inequalities \ref{MinUncCZ1}-\ref{MinUncCZ3} we find that:
\begin{eqnarray}
\label{KevinProd1a}
\Delta (x_1-x_2) \Delta (p_1+p_2+p_3) &\geq& \sum_{i}\eta_{i} \Delta (x_1-x_2)_{i} \Delta (p_1+p_2+p_3)_{i} + \sum_{j}\mu_{j} \Delta (x_1-x_2)_{j} \Delta (p_1+p_2+p_3)_{j} \\ \notag
&&+\sum_{k}\gamma_{k} \Delta (x_1-x_2)_{k} \Delta (p_1+p_2+p_3)_{k}+\sum_{l}\nu_{l} \Delta (x_1-x_2)_{l} \Delta (p_1+p_2+p_3)_{l}.
\end{eqnarray}

If the component of the state $ \rho_{i,12} $ is  entangled across particles $1$ and $2$, then the lowest possible value of $\Delta (x_1-x_2)_{i} \Delta (p_1+p_2+p_3)_{i}$ is zero.  The remaining three terms are for cases where particles $1$ and $2$ are on opposite sides of a separable cut; from equation \ref{FurusawaProd1} the minimum value of the uncertainty products are $\Delta (x_1-x_2)_{j} \Delta (p_1+p_2+p_3)_{j}= \Delta (x_1-x_2)_{k} \Delta (p_1+p_2+p_3)_{k}= \Delta (x_1-x_2)_{l} \Delta (p_1+p_2+p_3)_{l}=1$.  The uncertainty product in equation \ref{KevinProd1a} can then be written as:
\begin{eqnarray}
\label{KevinProd1}
\Delta (x_1-x_2) \Delta (p_1+p_2+p_3) &\geq& \sum_{i}\eta_{i}(0) + \left(\sum_{j}\mu_{j} + \sum_{k}\gamma_{k}+\sum_{l}\nu_{l}\right)(1), \notag \\
&\geq& \sum_{j}\mu_{j} + \sum_{k}\gamma_{k}+\sum_{l}\nu_{l}.
\end{eqnarray}

Following the same arguments it is possible to show that:
\begin{eqnarray}
\label{KevinProd2}
\Delta (x_1-x_3) \Delta (p_1+p_2+p_3) &\geq& \sum_{i}\eta_{i} + \sum_{k}\gamma_{k}+\sum_{l}\nu_{l}.\\
\label{KevinProd3}
\Delta (x_2-x_3) \Delta (p_1+p_2+p_3) &\geq& \sum_{i}\eta_{i} + \sum_{j}\mu_{j}+\sum_{l}\nu_{l}.
\end{eqnarray}

Adding inequalities \ref{KevinProd1} and \ref{KevinProd2} yields the sum inequality:
\begin{eqnarray}
\left[ \Delta (x_1-x_2) +\Delta (x_1-x_3)\right] \Delta (p_1+p_2+p_3) &\geq& \sum_{i} \eta_{i} + \sum_{j} \mu_{j}+2 \left(  \sum_{k}\gamma_{k} +\sum_{l}\nu_{l}\right).
\end{eqnarray}
The right-hand side is minimized by setting $\sum_{k}\gamma_{k}=\sum_{l}\nu_{l}=0$ giving:
\begin{eqnarray}
\label{sum1}
\left[ \Delta (x_1-x_2) +\Delta (x_1-x_3)\right] \Delta (p_1+p_2+p_3) &\geq& 1.
\end{eqnarray}
Similarly we find:
\begin{eqnarray}
\label{sum2}
\left[ \Delta (x_1-x_2) +\Delta (x_2-x_3)\right] \Delta (p_1+p_2+p_3) &\geq& 1,\\
\label{sum3}
\left[ \Delta (x_2-x_3) +\Delta (x_1-x_3) \right] \Delta (p_1+p_2+p_3) &\geq& 1.
\end{eqnarray}
Violating any of these sum inequalities \ref{sum1}-\ref{sum3} indicates that the state cannot be written with at most biseparable terms, and thus must be genuine tripartite entangled. 

It is also possible to add the three inequalities \ref{KevinProd1}-\ref{KevinProd3} together:
\begin{eqnarray}
\left[ \Delta (x_1-x_2)  +\Delta (x_1-x_3) + \Delta (x_2-x_3)\right] \Delta (p_1+p_2+p_3) &\geq& 2\left( \sum_{i} \eta_{i} + \sum_{j} \mu_{j}+ \sum_{k}\gamma_{k}  \right)+3 \left(\sum_{l}\nu_{l}\right),
\end{eqnarray}
which is minimized when $\sum_{l}\nu_{l}=0$ leading to:
\begin{eqnarray}
\label{sumall3}
\left[ \Delta (x_1-x_2)  +\Delta (x_1-x_3) + \Delta (x_2-x_3)\right] \Delta (p_1+p_2+p_3) &\geq& 2.
\end{eqnarray}
Violating this inequality also demonstrate the presence of genuine tripartite entanglement.

\subsection{Example states}

Consider the following three states of three different particles written in the position basis:
\begin{eqnarray}
\label{psi1}
\psi_{1}(x_{1},x_{2},x_{3})&=&\frac{1}{N_{1} }e^{-\left( \frac{x_{1}} { 2 \sigma_{1}}\right)^{2}} e^{-\left( \frac{x_{2}} { 2 \sigma_{2}}\right)^{2}} e^{-\left( \frac{x_{3}} { 2 \sigma_{3}}\right)^{2}} e^{-\left( \frac{x_{1}-x_{2}} { 2 \sigma_{c}}\right)^{2}} ,\\
\label{psi2}
\psi_{2}(x_{1},x_{2},x_{3}) &=& \frac{1}{N_{2} }e^{-\left( \frac{x_{1}} { 2 \sigma_{4}}\right)^{2}} e^{-\left( \frac{x_{2}} { 2 \sigma_{5}}\right)^{2}} e^{-\left( \frac{x_{3}} { 2 \sigma_{6}}\right)^{2}} e^{-\left( \frac{x_{1}-x_{3}} { 2 \sigma_{c}}\right)^{2}} ,\\
\label{psi3}
\psi_{3}(x_{1},x_{2},x_{3}) &=& \frac{1}{N_{3} }e^{-\left( \frac{x_{1}} { 2 \sigma_{7}}\right)^{2}} e^{-\left( \frac{x_{2}} { 2 \sigma_{8}}\right)^{2}} e^{-\left( \frac{x_{3}} { 2 \sigma_{9}}\right)^{2}} e^{-\left( \frac{x_{2}-x_{3}} { 2 \sigma_{c}}\right)^{2}} ,
\end{eqnarray}
where $N_{1}$, $N_{2}$, and $N_{3}$ are normalization constants and $\sigma_{1}$ through $\sigma_{6}$ are adjustable width parameters. The width $\sigma_{c}$ represents a correlation length for each of the states. When $\sigma_{c}\rightarrow0$  two of the particles in each of the states become perfectly entangled with one another.   The lower bound of the derived uncertainty relations \ref{sum1}-\ref{sumall3} are saturated by pure states of the form shown in equations \ref{psi1}-\ref{psi3}, and are therefore tight.

In general, full inseparability does not imply genuine tripartite entanglement. Consider the fully inseparable state composed of an equal mixture of $\psi_{1}$ and $\psi_{2}$ with $\sigma_{c}=0$, $\sigma_{2,3,5,6}=1$, and $\sigma_{1,4}\rightarrow \infty$. This mixed state, despite containing no component of genuine tripartite entanglement, simultaneously violate the van Loock and Furusawa inequalities \ref{FurusawaProd1} and \ref{FurusawaProd3}. However, when tested against the new inequality derived in \ref{sum1}, it reaches a minimum value of $\sqrt{2} > 1$; our inequality is able to discern that this state contains no genuine tripartite entanglement. A similar fully inseparable mixed state, composed of an equal mixture of $\psi_{1}$, $\psi_{2}$, and  $\psi_{3}$  with $\sigma_{c}=0$, $\sigma_{1,2,4,6,8,9}=1$, and $\sigma_{3,5,7}=1/\sqrt{2}$, is capable of simultaneously violating \emph{all three} of the van Loock and Furusawa inequalities despite not being genuinely tripartite entangled. When tested against the new inequality \ref{sumall3} it reaches a value of $\sqrt{6}>2$, demonstrating that it is not genuinely tripartite entangled.  

In the specific case of pure states,  full inseparability implies genuine tripartite entanglement. With the assumption of purity, both the van Loock and Furusawa inequalities as well as the ones derived here are capable of detecting genuine tripartite entanglement. For example, the pure state
\begin{eqnarray}
\label{psi4}
\psi_{4}(x_{1},x_{2},x_{3})&=&\frac{1}{N_{4} }e^{-\left( \frac{x_{1}} { 2 \sigma_{1}}\right)^{2}} e^{-\left( \frac{x_{2}} { 2 \sigma_{2}}\right)^{2}} e^{-\left( \frac{x_{3}} { 2 \sigma_{3}}\right)^{2}} e^{-\left( \frac{x_{1}-x_{2}} { 2 \sigma_{c}}\right)^{2}} e^{-\left( \frac{x_{1}-x_{3}} { 2 \sigma_{c}}\right)^{2}},
\end{eqnarray}
where $N_{4}$ is a normalization constant, exhibits perfect correlations between particles 1-3 when $\sigma_{c}\rightarrow0$ (for finite widths $\sigma_{1,2,3}$), and is both fully inseparable and genuinely tripartite entangled.  This pure state will violate the van Loock and Furusawa inequalities as well as the inequalities \ref{sum1}-\ref{sum3} and \ref{sumall3}. It is only in the more general case of mixed states that the relations derived by van Loock and Furusawa need to be extended in order to detect genuine tripartite entanglement.

\section{Alternate measurement of the photon timing uncertainty}
\begin{figure}
\centering
\includegraphics[width=5.5in]{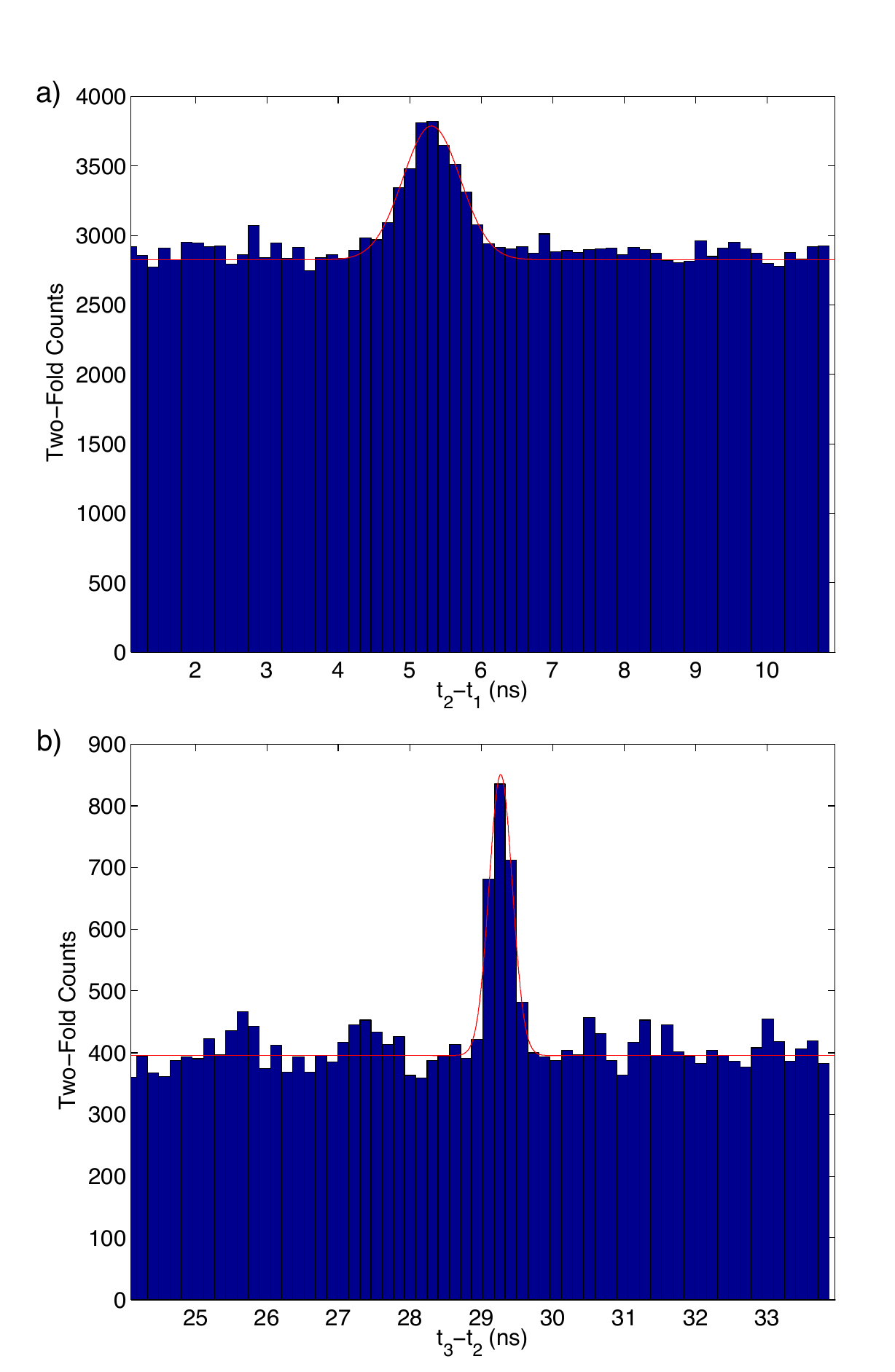}
\caption{Histograms of the arrival times between a) photons 1 and 2 and b) photons 2 and 3.  This was accomplished by examining the two-fold coincidences, as opposed to the three-fold coincidences, from the collected time tags.  From these histograms an uncertainty in the difference between the arrival times was found to be $\Delta(t_{2}-t_{1})=0.4\pm0.2$ ns and $\Delta(t_{3}-t_{2})=0.16\pm0.04$ ns which agrees with the integrated uncertainties measured in the main paper.}
\label{F-doublesHist}
\end{figure}

To verify that integrating over the third photon yields the correct two-photon timing histograms, the raw two-photon coincidence data from the measured time tags between photons 1 and 2 as well as  the coincidences between photons 2 and 3 were analyzed independently of detecting the other photon.  While the timing uncertainty between two detectors in Equations 2-4 are conditioned on the presence of a third photon, this measurement of the arrival times (shown in Figure \ref{F-doublesHist}) is different in that it is conditioned only on two photons.  From this data, it was found that $\Delta(t_{2}-t_{1})=0.4\pm0.2$ ns and $\Delta(t_{3}-t_{2})=0.16\pm0.04$ ns, which agree with the integrated values measured in the paper.  The reason for the larger error is due to the large number of accidental counts.  In the histogram between $t_{2}-t_{1}$ the large background is caused by the coincidences between dark counts on the free-running InGaAs/InP detector and the $10^{6}$ trigger photons per second on the Si detector.  The number of dark counts, and hence signal-to-noise ratio, is lower in the $t_{3}-t_{2}$ histogram; here the background is primarily due to accidental coincidences between dark counts between the two telecom detectors.  As the first telecom detector $t_{2}$ gated the second telecom detector $t_{3}$, our setup did not allow us to measure the two-fold coincidence histogram $t_{3}-t_{1}$ independent of photon 2.  

\clearpage
\section{Two photon energy-time entanglement}
\begin{figure}
\centering
\includegraphics[width=6in]{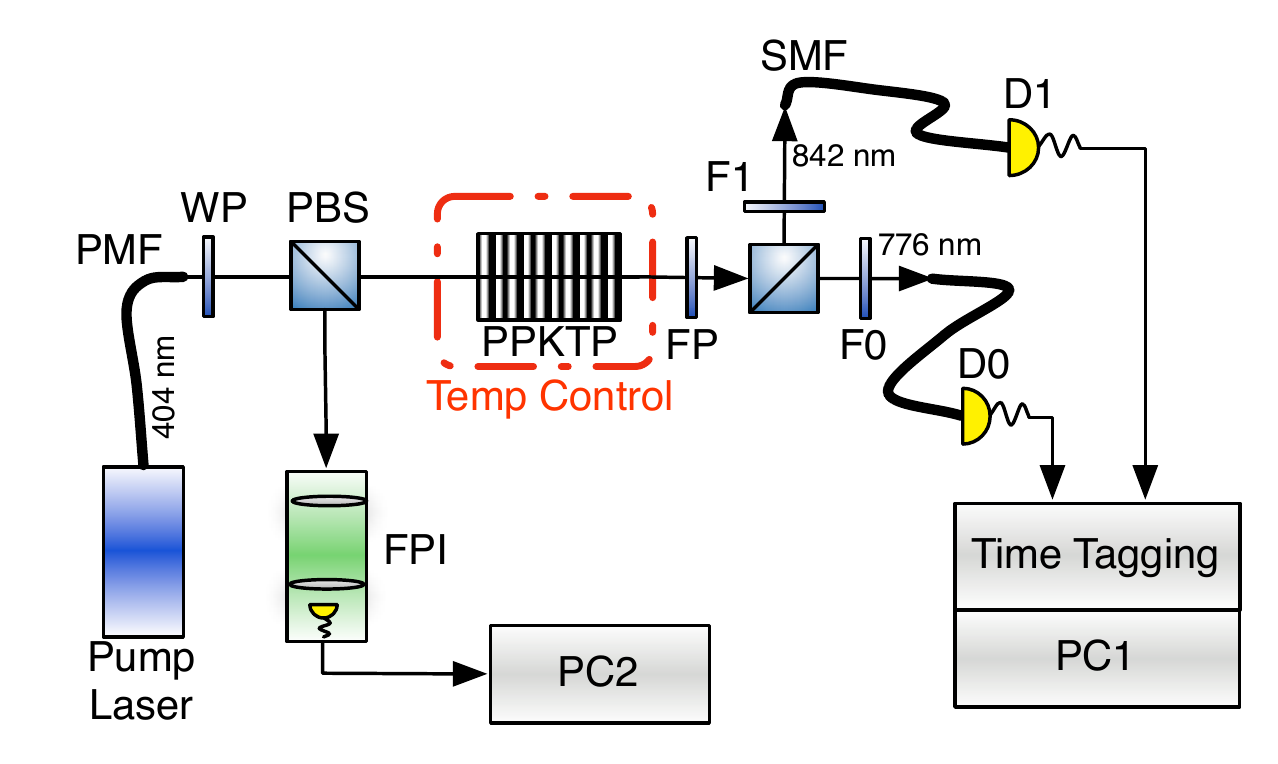}
\caption{Direct measurement of two-photon energy-time entanglement.  A narrowband pump laser at 404 nm is used to pump a periodically-poled Potassium Titanyl Phosphate (PPKTP) that downconverts into a pair of orthogonally polarized photons at 842 nm and 776 nm.  A filter (FP) removes the remaining pump.  A polarizing beamsplitter is used to separate the two photons, and narrowband filters, F0 and F1, are used to minimize stray light.  The photons at 842 nm and 776 nm are coupled into single-mode fibres and sent to the single-photon detectors  D0 and D1.  The signals from both detectors are sent to a time tagging unit, and a computer (PC1) is used to process coincidence events.  The spectrum of the 404 nm pump laser is continuously monitored throughout the run using a Fabry-Perot interferometer (FPI) controlled by a second computer (PC2). }
\label{F-Exp2}
\end{figure}

\begin{figure}
\centering
\includegraphics[width=6in]{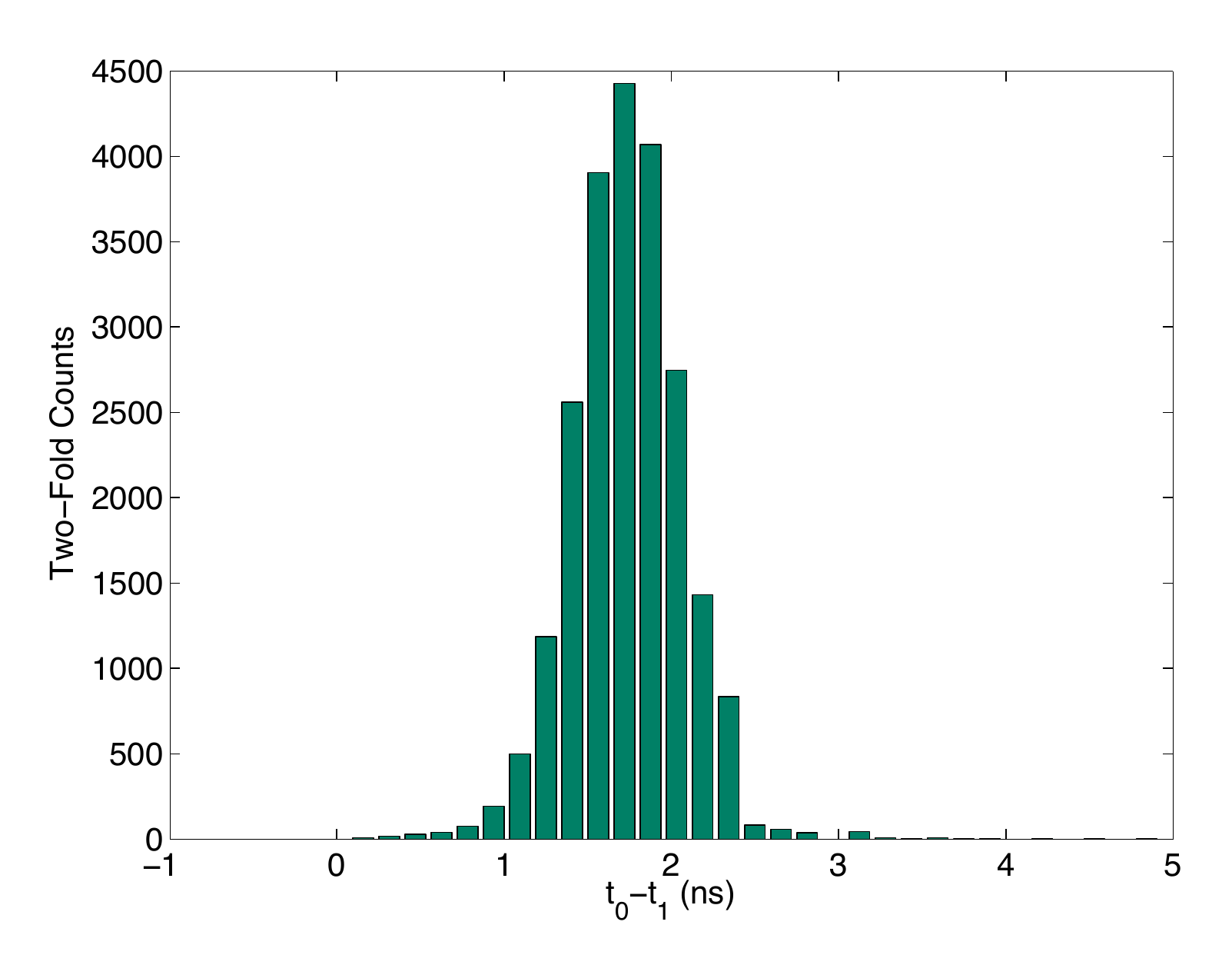}
\caption{Two-photon coincidence histogram of the difference in arrival times of the 842 nm and 776 nm photons from downconversion.  From this histogram, the uncertainty in the difference in arrival times was measured to be $\Delta (t_{0}-t_{1})=0.30\pm0.01$ ns}
\label{F-Doubles}
\end{figure}

In a separate experiment we directly measured for 1.6 s the two photon energy-time entanglement between the pair of daughter photons (at 842 nm and 776 nm) created at the first stage of our C-SPDC process as shown in Figure\ref{F-Exp2}.  Both daughter photons were sent to single-photon detectors.  The pump power was attenuated so that approximately 14,000 coincidences/s were detected.  A time tagging unit was used to store the arrival times of the photons, and a computer then sorted the time tags into extract coincidence events.    The histogram of arrival times of the two photons is shown in Figure \ref{F-Doubles}, and the uncertainty in the arrival times measured to be $\Delta (t_{0}-t_{1})=0.30\pm0.01$ ns.  Simultaneously the Fabry-Perot interferometer monitored the bandwidth of the pump, yielding a bandwidth of $\Delta \omega_{p}= 4.6 \pm 0.8$ MHz.

To show bipartite entanglement it is sufficient to violate the two party inequality $\Delta (\omega_{0}+\omega_{1}) \Delta (t_{0}-t_{1}) \geq 1$ \cite{vanLoock:2003hn}.  As energy is conserved in SPDC we can use the fact that $\omega_{0}+\omega_{1}=\omega_{p}$.  This leads to a time-bandwidth product of $\Delta (\omega_{0}+\omega_{1}) \Delta (t_{0}-t_{1})= 0.0014\pm 0.0002$ that violates the classical limit by more than 4000 standard deviations.

%%%%%%%%%%%%%%%%
%Experiment Figure Supp Info

%\bibliographystyle{naturemag}
%\bibliography{Energy-time}

 \end{document}